\documentclass[twocolumn,showpacs,aps,floatfix,superscriptaddress]{revtex4}
\usepackage{amsmath,amssymb,eucal,graphicx}
\usepackage{amsfonts}

\begin{document}
\title{The Simplest Piston Problem I: Elastic Collisions}
\author{Pablo I. Hurtado}
\email{phurtado@buphy.bu.edu}
\affiliation{Institute \emph{Carlos I} for Theoretical and Computational Physics, \\
Universidad de Granada, 18071 Granada, Spain}
\affiliation{Department of Physics, Boston University, Boston, Massachusetts 02215, USA}
\author{S.~Redner}
\email{redner@bu.edu}
\affiliation{Theoretical Division and Center for Nonlinear Studies, Los Alamos National Laboratory,
Los Alamos, New Mexico 87545, USA}
\altaffiliation{Permanent address: Department of Physics, Boston University, 
Boston, Massachusetts 02215, USA}

\begin{abstract}
  
  We study the dynamics of three elastic particles in a finite interval where
  two light particles are separated by a heavy ``piston''.  The piston
  undergoes surprisingly complex motion that is oscillatory at short time
  scales but seemingly chaotic at longer scales.  The piston also makes
  long-duration excursions close to the ends of the interval that stem from
  the breakdown of energy equipartition.  Many of these dynamical features
  can be understood by mapping the motion of three particles on the line onto
  the trajectory of an elastic billiard in a highly skewed tetrahedral
  region.  We exploit this picture to construct a qualitative random walk
  argument that predicts a power-law tail, with exponent $-3/2$, for the
  distribution of time intervals between successive piston crossings of the
  interval midpoint.  These predictions are verified by numerical
  simulations.
  
\end{abstract}
\maketitle

\section{INTRODUCTION}

A classic thermodynamics problem is the adiabatic ``piston'' \cite{C60},
where a gas-filled container is divided into two compartments by a heavy but
freely moving piston.  The piston is clamped in a specified position and the
gases in each compartment are prepared in distinct equilibrium states.  The
piston is then unclamped and the composite system evolves to a global
equilibrium.  This simple scenario leads to surprisingly complex dynamics
that are still incompletely understood, in both the cases where the two gases
are elastic \cite{L99,KVM00,GPL02,CLS02} and where they are inelastic
\cite{BRV05}.  In the elastic system, the piston moves quickly to establish
mechanical equilibrium where the pressures in each compartment are equal.
Subsequently, the piston develops oscillations that decay slowly as true
thermal equilibrium is achieved \cite{L99,KVM00,GPL02,CLS02}.  For the
inelastic system, there is a spontaneous symmetry breaking in which the gas
on one side of the piston gets compressed into a solid \cite{BRV05}.
Surprisingly, this process is not monotonic, but rather, the piston undergoes
oscillatory motion whose period grows exponentially with time.

\begin{figure}[b] 
 \vspace*{0.cm}
\includegraphics*[width=0.3\textwidth]{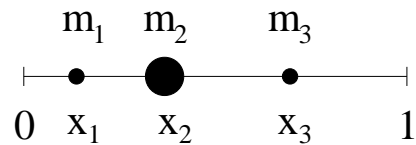}
\caption{The three-particle system---a piston and two light particles.
\label{system}}
\end{figure}

Given the complexities of these many-body problems, we instead investigate a
much simpler version (Fig.~\ref{system}): a three-particle system \cite{3p} in
the interval $0\leq x\leq 1$ consisting of two light particles of masses
$m_1=m_3=1$ that are separated by a heavy piston of mass $m_2\gg 1$.  All
interparticle collisions and collisions between particles and the ends of the
interval (henceforth termed walls) are elastic.  We will develop a simple
geometric approach and complementary numerical simulations to help understand
the complex dynamical features of this idealized system.  These results may
ultimately be useful for understanding the many-body piston problem.

An additional motivation to investigate the three-particle system is the
connection to the collective behavior in one-dimensional ($1D$) fluids. The
dimensional constraint induces strong interparticle correlations that lead to
anomalous transport properties.  For example, heat conductivity is generally
extremely large for $1D$ fluids, while mass diffusion is exceedingly slow
\cite{Livi}.  An example of a $1D$ fluid that exhibits such phenomenology is
a gas of point particles with alternating masses \cite{fluid}.  This fluid
can be viewed a collection of three-particle subsystems, each similar to our
idealized model.  We therefore anticipate that the dynamics of our three-particle
system can shed light on anomalous collective phenomena that arise in $1D$
fluids.

In the next section, we outline the basic phenomenology of the three-particle
system.  Then in Sec.~III, we map the trajectories of three particles on the line
onto an equivalent elastic billiard particle that moves within a
highly skewed tetrahedron, with the specular reflection whenever the billiard
hits the tetrahedron boundaries \cite{GZ,KT,T,G,R04}.  From this simple
geometrical mapping, we can understand many of the unusual dynamical
properties of the system, as will be discussed in Sec~IV.  Perhaps the most
unexpected feature is the long excursions of the piston close to the walls.
By the billiard equivalence, we will argue, in Sec.~V, that these long
excursions can be understood as the collision point of the billiard in the
tetrahedron undergoing a random walk.  We will thereby find that the
distribution of interval midpoint crossing times by the piston has a
power-law tail with exponent $-3/2$.

In an accompanying paper, we will consider the three-particle system when
collisions between the light particles and the walls are inelastic.
Surprisingly, much of the phenomenology of this idealized three-particle system
closely mirrors the complex dynamics that arises in the many-particle
inelastic piston problem \cite{BRV05}.

\begin{figure}[t!] 
  \vspace*{0.cm} \includegraphics*[width=0.45\textwidth]{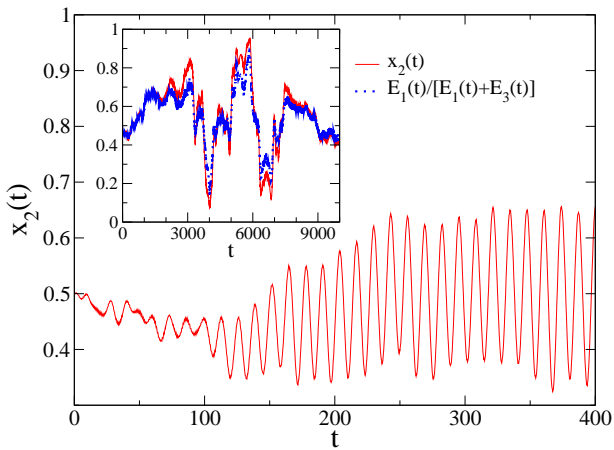}
\caption{(Color online) Trajectory of the piston for $t<400$ for $m_2=100$
  and initial positions $x_1=0.3$, $x_2=0.5$, and $x_3=0.7636$; 
  note the offset of the vertical axis.  Inset: Trajectory for $t<10^4$, as
  well as the estimate $x_2(t)\approx E_1(t)/[E_1(t)+E_3(t)]$ (see text).
  These data are averaged over a 400-point range for ease of visualization.
  \label{position}}
\end{figure}

\section{Piston Motion}
\label{motion}

To appreciate the basic phenomena, we show a typical piston trajectory
obtained numerically in Fig.~\ref{position} for the case $m_2=100$.  We focus
on the case $m_2=100$ because the overall simulation time scales as
$\sqrt{m_2}$ and there is already good scaling behavior for $m_2=100$.
However, we find qualitatively similar behavior for other values of $m_2\gg
1$.  For concreteness and simplicity, we also consider the initial condition
in which the light particles approach a stationary piston at $x_2=1/2$ with
velocities $v_1=+1$ and $v_3=-1$, so that the total momentum equals 0 while
the total energy $E=1$.  The initial particle positions of the light
particles are uniformly distributed in (0,1/2) and (1/2,1), respectively.
Again, we also studied different initial speeds for the light particles and
verified that our asymptotic results for the main quantities of interest do
not qualitatively depend on the initial conditions.

For the initial condition in which the light particles approach the piston
with velocities $v_1=+1$ and $v_3=-1$, we find that after a short transient
for $t\alt 100$, the piston settles into a quasiperiodic motion with period
of $T\approx 12$.  This time scale can be understood from simple arguments:
if there is energy equipartition, the piston would have energy 1/3 and speed
$\vert v_2\vert= \sqrt{2/3m_2}\approx 0.08$.  Equipartition also implies that
the typical spatial range of all three particles should be equal.  These two
features lead to a period of the order of 12, in agreement with the data.

In the limit $m_2 \to \infty$, it can be shown that the piston position obeys
(see Ref.~\cite{Sinai} and the Appendix)
\begin{equation}
\label{ode}
\frac{d^2x_2(t_s)}{dt_s^2} = \frac{A_1}{x_2^3(t_s)} - \frac{A_3}{[1-x_2(t_s)]^3}~,
\end{equation}
corresponding to a particle moving in an effective potential well $V_{\rm
  eff}(x)= \frac{1}{2}[A_1x^{-2} + A_3(1-x)^{-2}]$.  Here $t_s\equiv
t/\sqrt{m_2}$ is a \emph{slow} time variable that is a natural scale for the
piston motion, and $A_{1,3}$ are initial condition-dependent constants.  For
total energy $E=1$, we numerically determine from Eq.~(\ref{ode}) that the
oscillatory period in slow time coordinates is $T_s\approx 1.285$.  Thus a
piston with $m_2=100$ should oscillate with period $T=T_s\sqrt{m_2}\approx
12.85$, in excellent agreement with simulations (Fig.~\ref{position}).  Thus
this effective potential picture, which formally applies in the limit $m_2
\to \infty$, quantitatively accounts for the short-time oscillations of a
heavy but finite-mass piston.

\begin{figure}[t!] 
 \vspace*{0.cm}
\includegraphics*[width=0.45\textwidth]{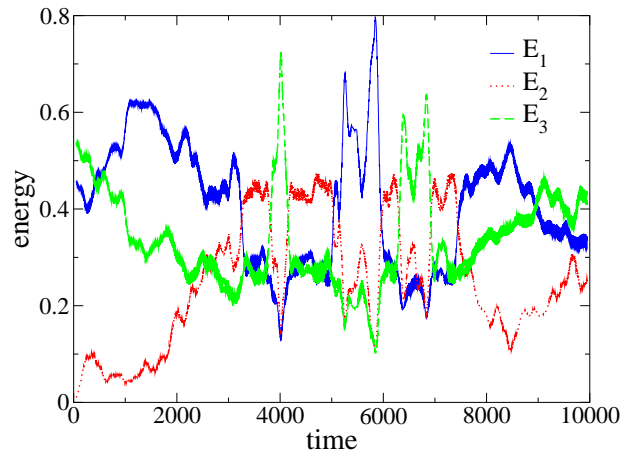}
\caption{(Color online) Particle energies as a function of time averaged over 
  a 400-point range.
  \label{energy}}
\end{figure}

For $t\agt 2000$, however, a considerably slower and much less predictable
large-amplitude modulation is superimposed on the quasiperiodic oscillations
(inset to Fig.~\ref{position}).  When $m_2=100$, the piston eventually
approaches to within of 0.05 of each wall.  These long-time extreme
excursions are reflected in the time dependence of the particle energies
(Fig.~\ref{energy}).  For $t>2000$, the piston energy fluctuates strongly and
is phase locked with $x_2(t)$ during the extreme excursions.  Notice also
that for $t<2000$ the piston energy is consistently below its average
long-time value, indicating the extent of the transient regime.

We may alternatively estimate the piston position by mechanical equilibrium
and basic thermodynamics.  We write $P_i\ell_i \propto E_i$, where $P_i$ and
$E_i$ are, respectively, the pressure and energy associated with particles
$i=1, 3$, and $\ell_i$ is the length available to particle $i$.  Assuming
mechanical equilibrium, $P_1=P_3$, and using $\ell_1=1-\ell_3=x_2$, we find
$x_2(t)= E_1(t)/[E_1(t)+E_3(t)]$.  This is very close to the numerical data
for $x_2(t)$ (inset to Fig.~\ref{position}); thus the piston excursions and
the large energy fluctuations away from equipartition are closely connected.

Finally, Eq.~(\ref{ode}) can be derived heuristically.  The equation of
motion for the piston is $m_2\dot v_2=F_1-F_3$, where the overdot denotes the time
derivative and $F_i$ is the force exerted on the piston by particle $i$.
Consider a time range large compared to the typical time between successive
light particle bounces, but small compared to the time for the piston to move
a unit distance.  Then $F_i\approx \Delta p_i/\Delta t_i$, where $\Delta
p_i=-\Delta v_i$ is the momentum change of the piston after a collision with
particle $i$, and $\Delta t_i$ is the time between successive bounces of the
particle with the piston.  When particle 1 with velocity $v_1$, collides with
the piston with velocity $v_2$, the outgoing velocity of the former is
$2v_2-v_1$ in the limit $m_2\gg 1$.  Since $v_2\sim {O}(m_2^{-1/2})$, we
have $\Delta p_1\approx 2v_1$.  Now $\Delta t_1\approx 2\ell_1/v_1$, where
$\ell_1=x_2$ is the length of the subinterval that contains particle 1.
Parallel results hold for collisions between the piston and particle 3.  Thus
\begin{eqnarray}
\label{v2}
m_2\dot v_2 =\frac{v_1^2}{x_2}- \frac{v_3^2}{1-x_2}\,.
\end{eqnarray}

To obtain $v_1$, note that after reflection from the left wall, particle 1
approaches the piston with velocity $v_1-2v_2$, so the net change in $v_1$
between successive collisions with the piston is $-2v_2$.  Thus the velocity
of particle 1 evolves according to $\dot v_1 \approx -2v_2/(2\ell_1/v_1)=
-v_1 \dot x_2/x_2$, with solution $v_1\propto 1/x_2$.  An analogous equation
holds for $v_3$.  Using these results in Eq.~(\ref{v2}) gives
Eq.~(\ref{ode}).

\section{Billiard Mapping}

To help understand the unusual features of the particle trajectories, it
proves useful to follow conventional practice \cite{GZ,KT,T,G,R04} and map
the three-particle system into an equivalent effective billiard.  To be general,
suppose that the particles have masses $m_1$, $m_2$, and $m_3$, are located
at $0\leq x_1(t)\leq x_2(t)\leq x_3(t)\leq 1$, and have velocities $v_1(t)$,
$v_2(t)$, and $v_3(t)$.  The trajectories of the three particles in the
interval are then equivalent to the trajectory $(x_1(t),x_2(t),x_3(t))$ of an
effective billiard particle in the three-dimensional domain defined by the
constraints $0\leq x_1\leq x_2\leq x_3\leq 1$.  For example, a collision
between particle 1 and the left wall corresponds to the billiard ball hitting
the boundary $x_1=0$, while a collision between particles 1 and 2 corresponds
to the billiard hitting the boundary $x_1=x_2$, etc.

\begin{figure}[t] 
 \vspace*{0.cm}
\includegraphics*[width=0.4\textwidth]{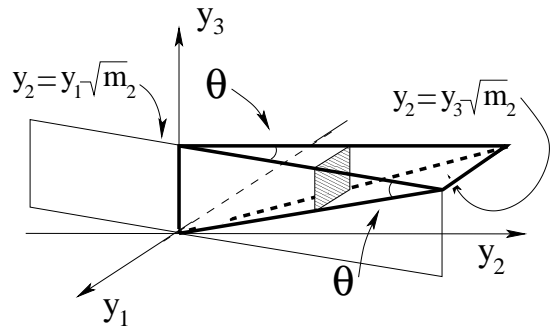}
\caption{The allowed tetrahedron (outlined by heavy lines) for the effective 
  billiard particle in $y_i$ coordinates.  The back and top planes of the
  tetrahedron are defined by $y_1=0$ and $y_3=1$, while the planes
  $y_2=y_1\sqrt{m_2}$ and $y_2=y_3\sqrt{m_2}$ correspond to 1-2 and 2-3
  collisions.
  \label{tetrahedron}}
\end{figure}

Unfortunately, momentum conservation shows that collisions between the
effective billiard and the boundaries of the domain are not specular.
Consequently, a naive analysis of successive billiard collisions becomes
prohibitively cumbersome.  However, a considerable simplification is achieved
by introducing the ``billiard'' coordinates \cite{GZ,KT,T,G,R04}
\begin{eqnarray}
\label{tform}
y_i=x_i\,\sqrt{m_i} \, , \quad w_i=v_i\, \sqrt{m_i} \, ,\qquad i=1,2,3\,.
\end{eqnarray}
In these coordinates, the constraints $x_1 \leq x_2$ and $x_2\leq x_3$ become
\begin{eqnarray*}
\label{constraint}
\sqrt{\frac{m_2}{m_1}} \;y_1\leq  y_2\, , \qquad  \sqrt{\frac{m_3}{m_2}} \;y_2\leq  y_3 \, ,
\end{eqnarray*}
while the constraints involving the walls are $y_1\geq 0$ and $y_3\leq
\sqrt{m_3}=1$.  As shown in Fig.~\ref{tetrahedron}, the allowed region for
the billiard is the interior of a highly skewed tetrahedron whose two acute
interior angles are given by $\theta= \tan^{-1}\sqrt{1/m_2}$.  While this
geometry may seem complicated at first sight, these coordinates ensure that
all billiard collisions with domain boundaries are specular \cite{T,G,R04},
and this feature greatly simplifies the problem.

We now exploit this billiard mapping to characterize the motion of the piston
in the original three-particle system.  For a zero-momentum initial condition,
the initial billiard trajectory lies within the shaded square
$y_2=\sqrt{m_2}/2$ (equivalent to $x_2=1/2$ in the interval) in
Fig.~\ref{tetrahedron}.  If the first collision is between the piston and
particle 1, the equivalent billiard first hits the front wall of the
tetrahedron.  Because of specularity, the billiard is reflected toward
increasing $y_2$.  Conversely, if the first collision is between the piston
and particle 3, the billiard first hits the bottom wall and the reflected
trajectory is toward decreasing $y_2$.

The opposite effects of successive 1-2 and 2-3 collisions lead to the
billiard persisting close to the shaded square $y_2=\sqrt{m_2}/2$.  However,
once the billiard develops a nonzero velocity in the $y_2$ direction, the
trajectory is unlikely to return to the initial square.  Subsequently, the
billiard bounces back and forth primarily along the $y_2$ direction in the
tetrahedron, corresponding to the quasiperiodic oscillations in the interval
shown in Fig.~\ref{position}.  At still longer times, the billiard motion
consists of unpredictable modulations that are superimposed on the
quasiperiodic oscillations.  The long-lived excursions of the piston near
one end of the interval correspond to the billiard remaining close to one of
the acute-angled ends of the tetrahedron in Fig.~\ref{tetrahedron}.

Another useful consequence of the billiard mapping is that we can also deduce
in simple terms the probability distribution $\pi_2(x)$ for finding the
piston at position $x_2=x$ in the interval, or equivalently the probability
for finding the billiard with coordinate $y_2=x_2\sqrt{m_2}\equiv z$.  This
distribution can also be found by using the microcanonical ensemble (see 
  e.g., the first paper in Ref.~\cite{3p}).  If the billiard covers the
tetrahedron equiprobably, then $\pi_2(x)$ would be proportional to the area
of the rectangle defined by the intersection of the plane $y_2=z$ and the
tetrahedron in Fig.~\ref{tetrahedron}.  This mixing property is believed to
occur in triangular billiards with irrational angles \cite{triangle} and also
in various three-dimensional billiard geometries \cite{Sinai3d}.  Given that
the angles of our tetrahedron generically are irrational except for
particular values of $m_2$, we expect that billiard trajectories in this
tetrahedron will also be mixing.

From this mixing hypothesis, $\pi_2(z)$ is simply proportional to the area of
the rectangle $y_2=z$ in the tetrahedron.  Now the length of the horizontal
side of the rectangle is proportional to $z/\sqrt{m_2}$, while the length of
the vertical side is proportional to $1-z/\sqrt{m_2}$.  Thus the rectangle
area is proportional to $z/\sqrt{m_2}(1-z/\sqrt{m_2})=x_2(1-x_2)$.
Normalization of this probability fixes the proportionality constant and we
thus obtain $\pi_2(x)=6x(1-x)$ for the probability that the piston is located
at $x$.  Similarly, the position distribution of the light particles is found
by computing the areas of the triangles defined by the intersection of the
planes $y_i=x\sqrt{m_i}$ ($i=1,3$) with the tetrahedron.  This leads to
$\pi_1(x)=3(1-x)^2$ and $\pi_3(x)=3x^2$.

We tested these predictions numerically and obtained excellent agreement
between the above theoretical expectations and the simulation results.
Notice that under the assumption of the billiard visiting all points in the
tetrahedron equiprobably, the probability of finding {\em any} particle at a
given position on the interval is a constant; that is, $\Pi(x)\equiv
\frac{1}{3}\sum_i \pi_i(x) = 1$.

\begin{figure}[t] 
  \vspace*{0.cm}
  \includegraphics*[width=0.45\textwidth]{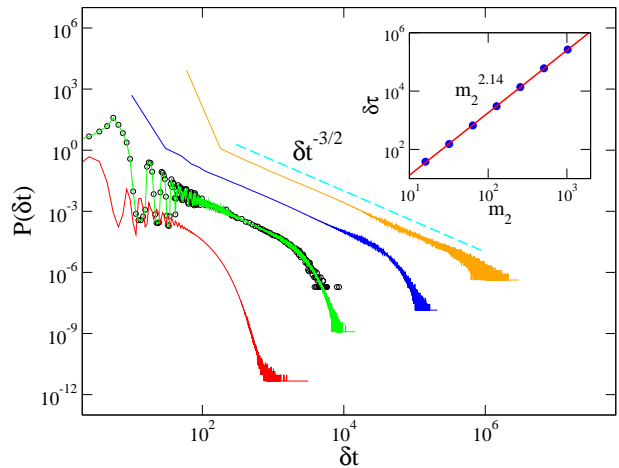}
  \caption{(Color online) The distribution of midpoint crossing time
    intervals $P(\delta t)$ for $m_2=2^n$, with $n= 4,6,8,$ and 10 on a
    double logarithmic scale (lower to upper curves) for zero initial
    momentum of the system.  For the case $m=64$, we also show as open
    circles the distribution for random initial conditions (positions and
    velocities) of the light particles. For visibility, each successive curve
    is shifted vertically upward by $10^{n-4}$.  Early-time oscillations do
    not appear for large $m_2$ because the histogram bin is larger than
    interpeak spacing in Fig.~\ref{tcrossing2}.  Inset: The cutoff
    $\delta\tau$ as a function of $m_2$.
  \label{tcrossing}}
\end{figure}

\section{Extreme Excursions}

To characterize the wanderings of the piston near the ends of the interval,
we study the probability distribution $P(\delta t)$ to have a time interval
$\delta t$ between successive midpoint crossings by the piston.  A midpoint
crossing corresponds to the equivalent billiard crossing the plane
$y_2=\sqrt{m_2}/2$.  As shown in Fig.~\ref{tcrossing}, $P(\delta t)$ decays
as the power law $(\delta t)^{-\mu}$ over a significant time range.  For the
case of $m_2=1024$, the data for $P(\delta t)$ versus $\delta t$ are quite
linear on a double logarithmic scale for $\delta t$ in the range $[1.8\times
10^2, 3.1\times10^5]$).  We measure the slope to be $\mu=1.5203\pm 0.0024$.
At longer times, the data have an exponential cutoff $P(\delta t)\sim
e^{-\delta t/\delta\tau}$, where $\delta\tau\sim m_2^{\lambda}$ with $\lambda
= 2.14\pm 0.01$ (inset to Fig.~\ref{tcrossing}).  Correspondingly, the
average time between crossings varies as $\langle\delta t\rangle \sim
m_2^{(2-\mu)\lambda}$.

\begin{figure}[t] 
 \vspace*{0.cm}
\includegraphics*[width=0.45\textwidth]{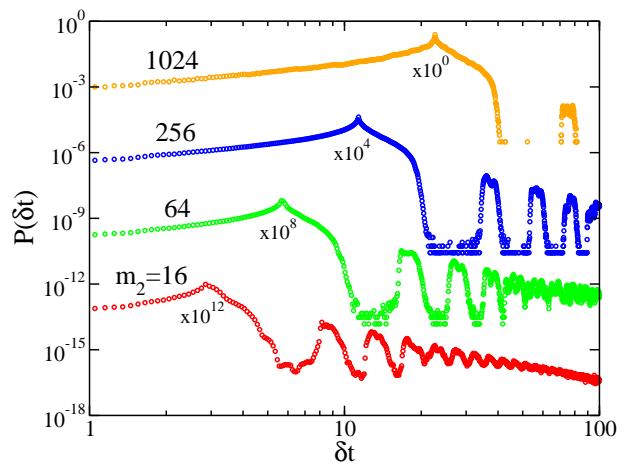}
\caption{(Color online) The short-time behavior of $P(\delta t)$ for 
  different $m_2$. 
\label{tcrossing2}}
\end{figure}

\begin{figure}[t] 
 \vspace*{0.cm}
\includegraphics*[width=0.45\textwidth]{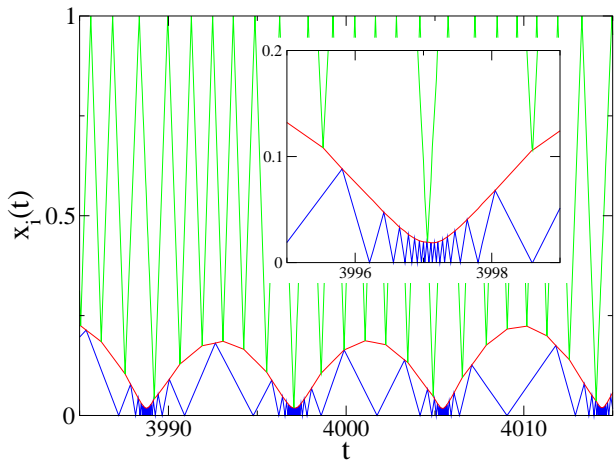}
\caption{(Color online) Positions of the three particles for the system of 
  Fig.~\ref{position} near $t= 4000$.  Inset: Finer detail near $t=3997$.
\label{micro}}
\end{figure}

From the relation $x_2(t) \approx E_1(t)/[E_1(t)+E_3(t)]$ derived in
Sec.~\ref{motion}, the piston crosses the midpoint whenever $E_1(t)=E_3(t)$;
thus $P(\delta t)$ can also be interpreted as the probability that the
inequality $E_1(t)\ne E_3(t)$ persists for a time $\delta t$.  This long-time
persistence of energy asymmetry is in agreement with previous simulations of
$1D$ binary fluids \cite{fluid}, in which light particles were reported to
trap energy and release it very slowly.  

The early-time sequence of peaks in $P(\delta t)$ is simply related to the
half-period of the short-time piston oscillations (and its resonances),
$\frac{1}{2}T= \frac{1}{2}T_s\sqrt{m_2}$, where $T_s\approx 1.285$ is the
slow-time period associated with a particle in the effective potential of
Eq.~(\ref{ode}).  Thus the first peak of $P(\delta t)$ should be at $\delta
t\approx 2.6$, 5.1, 10.3, and 20.6 for $m_2=16$, 64, 256, and 1024,
respectively, very close to the results in Fig.~\ref{tcrossing2}.

To understand the long-time power-law in $P(\delta t)$, we show in
Fig.~\ref{micro} the particle trajectories from Fig.~\ref{position} during
the extreme excursion near $t\approx 4000$.  This excursion is driven by a
sequence of nearly periodic oscillations due to precisely orchestrated
correlated motion of the lighter particles.  Consider first the collisions
between particle 1 and the piston when the latter moves toward $x=0$.  There
is a violent series of ``rattling'' collisions as the piston first approaches
$x=0$ and ultimately is reflected \cite{S,R04}.  In the limit $m_2\gg 1$,
these rattling collisions are equivalent to the piston having a nearly
elastic reflection from the wall.  After this rattling collision, the piston
is met by particle 3 whose momentum is of a similar magnitude, but opposite
to that of the piston.  Thus after a few collisions between the piston and
particle 3 (seven such collisions in Fig.~\ref{micro}) the piston is reflected
back toward $x=0$, where the rattling between particle 1 and the piston
recurs.

\section{Effective Random Walk Description}

To determine $P(\delta t)$ from this descriptive account of the rattling
collisions, we consider a reduced problem in which the fastest degrees of
freedom associated with particle 1 are integrated out.  As we shall show, the
slower degree of freedom associated with the piston can then be described
qualitatively by an effective random walk.  This then allows us to deduce the
statistics of the time between successive piston crossings of the interval
midpoint.

For the piston to persist near the left wall, the collisions between the
piston and particle 3 must be close to periodic.  A deviation from
periodicity occurs because the net effect of the rattling between particle 1
and the piston is a slightly inelastic collision.  We now estimate the
departure from elasticity in these rattling collisions and we then use this
result to estimate the duration of the resonance between the piston and
particle 3.

\begin{figure}[t] 
 \vspace*{0.cm}
\includegraphics*[width=0.375\textwidth]{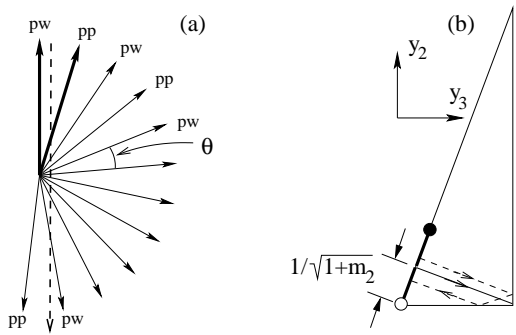}
\caption{(a) The wedge that represents collisions between particle 1, the
  piston, and the left wall in billiard coordinates (thick lines).  Lighter
  lines show the periodic extension of wedge. The dashed straight line is
  periodic extension of the billiard trajectory.  (b) The billiard after
  projection from the tetrahedron onto the $y_2$-$y_3$ triangle.  Shown is
  the periodic trajectory when the piston and particle 3 have equal and
  opposite momenta and meet at $x=1/(1+m_2)$.  A two-cycle arises (dashed) when
  the collision point deviates within the thick segment of the hypotenuse
  while the particle momenta remain equal and opposite.
\label{fan-triangle23}}
\end{figure}

In billiard coordinates, the rattling collision can be represented as the
effective billiard entering a narrow wedge of opening angle
$\theta=\tan^{-1}(1/\sqrt{m_2})$ [Fig.~\ref{fan-triangle23}(a)] that is the
projection of the tetrahedron onto the $y_1$-$y_2$ plane.  Because each
collision of the billiard with the wedge is specular, the ensuing rattling
sequence is equivalent to a straight trajectory in the periodic extension of
the wedge [Fig.~\ref{fan-triangle23}(a)].  Each collision is alternately
particle-particle and particle-wall, so that the identity of
periodically extended barriers alternates between $pp$ and $pw$.  The
rattling sequence ends when the billiard trajectory no longer crosses a wedge
boundary.  The crucial point is that the final billiard velocity vector
deviates by no more than an angle $\theta$ with respect to the two rays that
define the last wedge.

Suppose that the initial velocity vector is $\vec v^{(i)}\equiv (v_1,v_2)=
(0,-1)$, corresponding to $\vec w^{(i)}\equiv (w_1,w_2)= (0,-\sqrt{m_2})$.
If the final billiard trajectory is parallel to a $pw$ boundary in
Fig.~\ref{fan-triangle23}(a), then $\vec v^{(f)}$ is $(0,+1)$.  This
corresponds to a rattling sequence in which particle 1 begins and ends at
rest and the piston is elastically reflected.  Conversely, if the final
trajectory is parallel to a $pp$ boundary, then 
\begin{equation}
\vec w_f=\sqrt{m_2/(1+m_2)} \, (-1,-\sqrt{m_2}) \nonumber
\end{equation}
(note that $w_i^2=w_f^2$).
Translating to original coordinates, the minimum final piston speed is
$\sqrt{m_2/(1+m_2)}$.  Therefore rattling collisions lead to a final
piston velocity that lies within the narrow range $(1-1/2m_2,1)$ for
$m_2\gg 1$.

From this deviation from elasticity, we determine the time needed to disrupt
the resonance between the piston and particle 3.  Consider the two-particle
system consisting of the piston and particle 3 with initial velocities
$(v_2,v_3)=(1/m_2,-1)$ and with $x_2=x_3=x=1/(1+m_2)$.  This resonant
starting state ensures that the two particles hit the opposite ends of the
interval simultaneously and then meet again at $x=1/(1+m_2)$ with
$(v_2,v_3)=(1/m_2,-1)$ when all collisions are elastic.  In billiard
coordinates this periodic motion translates to a singular trajectory in the
projection of the tetrahedron onto the $y_2$-$y_3$ plane
[Fig.~\ref{fan-triangle23}(b)].  A 2-3 collision corresponds to the billiard
hitting the hypotenuse of the resulting triangle perpendicularly and the
simultaneous collision of the particles with the interval ends corresponds to
the billiard hitting at the right-angle corner of the triangle.  If the
initial position slightly deviates from $x=1/(1+m_2)$ while the initial
velocities are on resonance, then the subsequent motion is simply a two-cycle,
as indicated in the figure.

Now consider the influence of particle 1 on this resonance.  Due to the
slight inelasticity of the effective collision between the piston and the
left wall, the speed of the reflected piston changes stochastically by the
order of $m_2^{-1}$.  Consequently, the 2-3 collision point shifts from
$x=1/(1+m_2)$ to $x=1/(1+m_2)\pm{O}(1/m_2^2)$.  In billiard coordinates,
the return trajectory in Fig.~\ref{fan-triangle23}(b) is not exactly parallel
to the initial trajectory and the collision point on the hypotenuse moves
stochastically by an amount of the order of $m_2^{-3/2}$.  When this
collision point moves outside the thick line in Fig.~\ref{fan-triangle23},
the resonance between the piston and particle 3 terminates and the piston
crosses the interval midpoint shortly thereafter.

Thus the collision point on the hypotenuse undergoes a random walk on an
interval of length ${O}(m_2^{-1/2})$ with one end absorbing (open circle
in Fig.~\ref{fan-triangle23}) and the other end reflecting (solid dot).  The
probability that the billiard remains in this interval up to time $\delta t$
therefore scales as $P(\delta t) \sim \delta t^{-3/2}$ until an exponential
cutoff because of the finiteness of the interval \cite{Sidbook}.  The cutoff
time should be $L^2/D$, where $L\sim {O}(m_2^{-1/2})$ is the interval
length and $D\propto [{O}(m_2^{-3/2})]^2$ is the diffusion coefficient
associated with individual random-walk steps of length $m_2^{-3/2}$.  This
leads to a cutoff time $\delta\tau\sim m_2^2$, consistent with the simulation
result $\delta \tau\sim m_2^{2.14}$ shown in Fig.~\ref{tcrossing}.

\section{Discussion}

We introduced a toy version of the classic piston problem in which a massive
particle (the piston) separates a finite interval into two compartments, each
containing a single light particle.  In spite of its simplicity, the dynamics
of this three-particle system is surprisingly rich.  When all collisions are
elastic, the piston undergoes complex motion, with short-time quasiperiodic
behavior and seemingly chaotic behavior at long times.  The early-time can be
understood in terms of the piston oscillating in an effective potential well
$V_{\rm eff}=[A_1x^{-2}+A_3(1-x)^{-2}]$.  To understand the long-time
behavior, we mapped the motion of three particles in the interval onto that
of an effective equivalent billiard in a tetrahedral domain.  We then used
geometric methods that help explain some of the anomalous dynamical features
of the piston.

At long times, the piston moves in an apparently unpredictable fashion, with
long-lived excursions close to the ends of the interval during which large
departures from energy equipartition occur.  We quantified these extreme
excursions by studying the distribution of time intervals for the piston to
cross the interval midpoint.  This distribution has a power-law decay over a
wide time range, with exponent $-3/2$.  We argued that this phenomenon can be
recast as a first-passage problem of a random walk within a finite interval,
leading naturally to the above exponent value.

Although individual trajectories of the piston seem unpredictable, average
properties are considerably simpler.  We found a simple form for the
probability distribution $\pi_i(x)$ for finding particle $i$ at position $x$;
namely, $\pi_1(x)=3(1-x)^2$, $\pi_2(x)=6x(1-x)$, and $\pi_3(x)=3x^2$.  These
forms are a direct consequence of the billiard trajectory being mixing in the
tetrahedron.

The three-particle system studied in this work is clearly oversimplified to
faithfully model a many-particle piston system in three dimensions.
Nevertheless, the methods developed here may prove useful in understanding
few-particle elastic or granular systems and may help suggest new approaches to
deal with many-particle systems in higher dimensions.

\acknowledgments{We thank R. Brito for collaboration during the initial
  stages of this project.  SR thanks NSF grant No. DMR0227670 (BU) and DOE grant
  No. W-7405-ENG-36 (LANL) for financial support.  PIH acknowledges support from
  Spanish MEC, and thanks LANL and CNLS for hospitality of during part of
  this project.}

\appendix*

\section{Effective Potential for an Infinite-Mass Piston}

Here we show that Eq.~(\ref{ode}) governs the piston motion in the limit $m_2
\to \infty$ by specializing the general result of Sinai \cite{Sinai} to the
three-particle elastic system in the interval.  We assume a initially at
$x_2(0)=1/2$ with $v_2(0)=0$, and unit-mass particles 1 and 3
starting at random positions to the left and right of the piston,
respectively, with velocities $v_1(0)=+1$ and $v_3(0)=-1$.  Energy
conservation implies that $\vert v_2(t)\vert<\sqrt{2/m_2}$.  It is then
natural to define a slow time variable, $t_s=t/\sqrt{m_2}$, such that the
piston velocity is ${O}(1)$ in this time scale.

Consider an infinitesimal slow time interval $[t_s,t_s+\delta]$ during which
the piston moves a distance ${O}(\delta)$, while the number of 1-2 and
2-3 collisions is ${O}(\sqrt{m_2})$.  Let $k$ index each piston
collision; we define this collision index to run from $k_-+1$ to $k_+$ in
$[t_s,t_s+\delta]$.  The total number of collisions experienced by the piston
in this time range is $N=k_+-k_-$.  The particle velocities just before each
collision with the piston are given by
\begin{eqnarray}
v_2(k) & =&  (1-\epsilon)  v_2(k-1) + \epsilon  v_i(k-1),  \label{dyn1}\\ 
v_i(k) & =&   (\epsilon-1)  v_i(k-1) + \alpha  v_2(k-1),  \label{dyn2}
\end{eqnarray}
where $\epsilon=2/(1+m_2)$, $\alpha=\epsilon m_2$, $i=1,3$, and $v(k)$ is a
particle velocity just before the $(k+1)\textrm{st}$ piston collision.  For
large piston mass recollisions do not occur, that is, light particles always
hit a boundary before colliding again with the piston.  Therefore, $v_1(k)>0$
and $v_3(k)<0$ $\forall \ k\in[k_-,k_+-1]$.

Next we iterate the first term in Eq.~(\ref{dyn1}) to write $v_2(k_+)\equiv
v_2(t_s+\delta)$ in terms of $v_2(k_-)\equiv v_2(t_s)$.  Let $n_{12}$ and
$n_{23}$ be the number of 1-2 and 2-3 collisions in the sequence
$k_-+1,\ldots, k_+$, respectively, with $N=n_{12}+n_{23}$.  For $i\in
[1,n_{12}]$, we define $c_1(i)=k$ if and only if the $i{\rm th}$ 1-2 collision
corresponds to the $k{\rm th}$ collision in $k_-+1,\ldots, k_+$, so that
$c_1(i)\in[1,N]$ and similarly for $c_3(j)$, with $j\in[1,n_{23}]$.  With
these definitions, (\ref{dyn1}) gives
\begin{eqnarray}
v_2(k_+) & = & (1-\epsilon)^N v_2(k_-) \label{iter} \\
& + &  \epsilon \sum_{i=1}^{n_{12}}(1-\epsilon)^{N-c_1(i)} ~ v_1[k_-+c_1(i)-1] \nonumber \\
& + &  \epsilon \sum_{i=1}^{n_{23}}(1-\epsilon)^{N-c_3(i)} ~ v_3[k_-+c_3(i)-1]. \nonumber
\end{eqnarray}
The piston velocity in the slow time variable is $w_2(t_s)\equiv
\frac{dx_2}{dt_s}=\sqrt{m_2} ~ v_2 (t_s)$.  To derive a closed equation for
$w_2(t_s)$, we first take the limit $m_2 \to \infty$ and then $\delta \to 0$.
Using the definition of $w_2(t_s)$ and Eq.~(\ref{iter}), we find
\begin{equation}
\begin{split}
 w_2&(t_s+\delta) - w_2(t_s) = \big[(1-\epsilon)^N-1 \big]w_2(t_s)  \\
+& \epsilon\sqrt{m_2}\Big\{ \sum_{i=1}^{n_{12}}\big[(1-\epsilon)^{N-c_1(i)}-1\big] ~ 
v_1[k_-+c_1(i)-1] \\
+& \sum_{j=1}^{n_{23}}\big[(1-\epsilon)^{N-c_3(j)}-1\big] ~ v_3[k_-+c_3(j)-1]   \\
+& \sum_{i=1}^{n_{12}} v_1[k_-+c_1(i)-1] +  \sum_{j=1}^{n_{23}} v_3[k_-+c_3(j)-1] \Big\}\,. 
\end{split}
\end{equation}
We expand this expression for $m_2 \to \infty$, taking into account that
$\epsilon\sim {O}(m_2^{-1})$ and $n_{1,3}\sim {O}(\sqrt{m_2})$, to
obtain
\begin{equation}
\label{itera2}
\begin{split}
 w_2(t_s+\delta) - &w_2(t_s) = \epsilon\sqrt{m_2}\Big\{ \sum_{i=1}^{n_{12}} 
v_1[k_-+c_1(i)-1]  \\
 + &  \sum_{j=1}^{n_{23}} v_3[k_-+c_3(j)-1] \Big\} + {O}\left(\frac{1}{\sqrt{m_2}}\right).
\end{split}
\end{equation}
Because the large piston mass causes the light particle velocities to change
only slightly in the slow time interval $[t_s,t_s+\delta]$, we can write
\begin{equation}
\sum_{k=1}^{n_i} v_i[k_-+c_i(k)-1] \approx n_i v_i(t_s)
\label{approxv13}
\end{equation} 
for $i=1,3$, with $v_i(t_s)\equiv v_i(k_-)$, and where correction terms
vanish as $\delta \to 0$ \cite{Sinai}.  Within this approximation of nearly
constant light-particle velocities, the unscaled time intervals between
successive 1-2 and 2-3 collisions are $2x_2(t_s)v_1^{-1}(t_s)$ and
$-2[1-x_2(t_s)]v_3^{-1}(t_s)$, respectively.  Thus
\begin{eqnarray}
n_{12} \approx \frac{v_1(t_s)\sqrt{m_2}\,\delta }{2x_2(t_s)} \, , \qquad 
n_{23} \approx \frac{-v_3(t_s)\sqrt{m_2}\,\delta }{2\big(1-x_2(t_s)\big)} \, , 
\nonumber
\label{ncol3}
\end{eqnarray} 
with $v_1(t_s)>0$ while $v_3(t_s)<0$.  Using these results in
Eq.~(\ref{itera2}), we obtain, in the asymptotic limit,
\begin{equation}
\frac{dw_2(t_s)}{dt_s} = \frac{v_1^2(t_s)}{x_2(t_s)} - 
\frac{v_3^2(t_s)}{[1-x_2(t_s)]} ~ .
\label{odevel}
\end{equation}

We now derive the equation of motion for $v_i(t_s)$, $i=1,3$.  Here we
consider only particle 1, since the derivation for particle 3 is analogous.
Let us introduce a new index $q\in[1,n_{12}]$, such that $v_1(q)$ and
$v_2(q)$ are the velocities of particle 1 and the piston just before the
$(q+1){\rm st}$ 1-2 collision (notice a subtle difference from the previous
notation; between the $q{\rm th}$ and the $(q+1){\rm st}$ 1-2 collisions,
the piston may collide one or more times with particle 3).  From
Eq.~(\ref{dyn2}) we have
\begin{equation} 
v_1(q+1) = (1-\epsilon) ~ v_1(q) - \alpha ~ v_2(q)\,. \nonumber
\end{equation}
Notice the extra minus sign in this equation compared to Eq.~(\ref{dyn2}) to
account for the reflection of the light particle off the wall.  Iterating
this equation, we find
\begin{equation}
v_1(n_{12}) = (1-\epsilon)^{n_{12}} ~ v_1(0) - \alpha \sum_{q=0}^{n_{12}-1} 
(1-\epsilon)^{n_{12}-q-1} ~ v_2(q), \nonumber
\end{equation}
where now $v_1(n_{12})\equiv v_1(t_s+\delta)$ and $v_1(0)\equiv v_1(t_s)$.
Taking the limit $m_2 \to \infty$ now yields
\begin{equation}
v_1(t_s+\delta) - v_1(t_s) = -2 \sum_{q=0}^{n_{12}-1} v_2(q) + 
{O}\left(\frac{1}{\sqrt{m_2}}\right)\,. \nonumber
\label{itera3}
\end{equation}
We now write $v_2(q)$ as $v_2(0) + \sum_{k=1}^q \big[v_2(k)-v_2(k-1) \big]$.
Therefore,
\begin{equation}
  \sum_{q=0}^{n_{12}-1} v_2(q) = n_{12} v_2(0) + \sum_{q=0}^{n_{12}-1}\sum_{k=1}^q 
  \big[v_2(k)-v_2(k-1) \big].
\label{doublesum}
\end{equation}
The double sum in Eq. (\ref{doublesum}) can be demonstrated to be ${\cal
  O}(\delta^2)$ \cite{Sinai}, so it is negligible in the $\delta \to 0$
limit.  Therefore
\begin{equation}
v_1(t_s+\delta) - v_1(t_s) = -\frac{\delta v_1(t_s)\sqrt{m_2}}{x_2(t_s)}v_2(t_s) + 
{O}(\delta^2)\,. \nonumber
\end{equation}
Finally, for $m_2 \to \infty$ and $\delta \to 0$ we find
\begin{eqnarray}
\label{odev}
\frac{dv_1(t_s)}{dt_s} &=& - \frac{v_1(t_s)}{x_2(t_s)} \frac{dx_2}{dt_s}\,,\nonumber \\
\frac{dv_3(t_s)}{dt_s} &=& \frac{v_3(t_s)}{1-x_2(t_s)} \frac{dx_2}{dt_s}\,.
\end{eqnarray}
Integrating these equations yields $v_1(t_s)=B_1/x_2(t_s)$ and
$v_3(t_s)=B_3/\big[1-x_2(t_s)\big]$, with $B_{1,3}$ constants which depend on
the initial condition.  Using these solutions in Eq.~(\ref{odevel}) we
finally arrive to the piston equation of motion given in Eq.~(\ref{ode}).

\end{document}